\documentclass[conference, twocolumn]{IEEEtran}
\pdfoutput=1

\usepackage{color}

\IEEEoverridecommandlockouts
\usepackage{multirow}
\usepackage{hhline}
\usepackage{ifpdf}
\ifpdf
  \usepackage[pdftex]{graphicx} 
  \usepackage{epstopdf}         
\else
  \usepackage{graphicx}         
\fi

\usepackage{cite}
\usepackage[cmex10]{amsmath}
\usepackage{stfloats}
\usepackage{subcaption}
\usepackage{xcolor}
\usepackage{soul}
\usepackage{bbold}
\usepackage{amssymb}
\usepackage{booktabs}

\renewcommand\hl[1]{#1} 
\renewcommand\st[1]{} 

\begin{document}

\title{Deep Modulation (Deepmod): A Self-Taught PHY Layer for Resilient Digital Communications}

\author{Adam~Anderson,~\IEEEmembership{Senior Member,~IEEE}, Steven R.~Young,~\IEEEmembership{Member,~IEEE}, \\F. Kyle Reed,~\IEEEmembership{Student Member,~IEEE}, and Jason~M.~Vann
\thanks{A. Anderson is with Oak Ridge National Laboratory and joint faculty with the Department of Electrical and Computer Engineering, Tennessee Tech University, Cookeville, TN, 38505 USA e-mail: aanderson@tntech.edu.}
\thanks{S. Young, F. Reed, and J. M. Vann are with Oak Ridge National Laboratory, Oak Ridge, Tennessee email: [youngsr, reedfk, vannjm]@ornl.gov}
\thanks{This Research was sponsored by the Laboratory Directed Research and Development Program of Oak Ridge National Laboratory, managed by UT-Battelle, LLC, for the U.S. Department of Energy. This manuscript has been authored by UT-Battelle, LLC, under Contract No. DE-AC0500OR22725 with the U.S. Department of Energy. The United States Government retains and the publisher, by accepting the article for publication, acknowledges that the United States Government retains a nonexclusive, paid-up, irrevocable, worldwide license to publish or reproduce the published form of this manuscript, or allow others to do so, for the United States Government purposes. The Department of Energy will provide public access to these results of federally sponsored research in accordance with the DOE Public Access Plan(http://energy.gov/downloads/doe-public-access-plan).The authors are with the Oak Ridge National Laboratory, Oak Ridge, TN 37831 USA.
}}

\maketitle
\pagestyle{empty}
\thispagestyle{empty}

\begin{abstract}
Traditional physical (PHY) layer protocols contain chains of signal processing blocks that have been mathematically optimized to transmit information bits efficiently over noisy channels. Unfortunately, this same optimality encourages ubiquity in wireless communication technology and enhances the potential for catastrophic cyber or physical attacks due to prolific knowledge of underlying physical layers. Additionally, optimal signal processing for one channel medium may not work for another without significant changes in the software protocol. Any truly resilient communications protocol must be capable of immediate redeployment to meet quality of service (QoS) demands in a wide variety of possible channel media. Contrary to many traditional approaches which use immutable man-made signal processing blocks, this work proposes generating real-time blocks {\it ad hoc} through a machine learning framework, so-called deepmod, that is only relevant to the particular channel medium being used. With this approach, traditional signal processing blocks are replaced with machine learning graphs which are trained, used, and discarded as needed. Our experiments show that deepmod, using the same machine intelligence, converges to viable communication links over vastly different channels including: radio frequency (RF), powerline communications (PLC), and acoustic channels.
\end{abstract}


\section{Introduction}
\IEEEPARstart{M}{odern} digital communications systems are rooted in networks of basic point-to-point (P2P) enabled devices. To ensure performance of these systems, various quality of service (QoS) metrics must be addressed to ensure user satisfaction. These QoS metrics range from security (e.g. low probability of detection, intercept, and/or exploitation [LPI, LPD, LPE]) \cite{2010Cirincione} in tactical networks to latency \cite{2011O'Melia,2010Cottle}, assurance, and throughput \cite{2016Nightingale} in more traditional networks; however, QoS can quickly degrade in unknown or attacked channels. A fully resilient P2P link would need to use dynamic signal processing capabilities coupled with autonomous machine intelligence to overcome drastic changes in the channel.  This approach can be achieved with software-defined radios (SDRs) and machine learning (ML) and is intriguing in that it addresses the developing requirements of dynamic channels while increasing diversity and resilience. SDR and ML have seen increased use in the development of modern digital communications schemes including machine modems. The versatility of SDRs \cite{2014Goeller} to reconfigure radio links adaptively is used to increase the testing and development of digital communications. ML - the composing of machines to continuously improve (or learn) with experience \cite{mitchell1997machine} - has been applied across a broad scope of scientific and technological fields.  

ML has been shown to be influential in signal processing applications such as computer and machine vision\cite{2017Kuang,2016Dhivyaprabha,2016Allodi}, anomaly detection\cite{2016Valdes,2016Mohd,2016Murphree}, and natural language processing\cite{2014Kandasamy,2015Pollettini,2015Lakhanpal}. In traditional systems, a ``machine'' is built as a solution to a mathematical system model. However, many practical applications are exceedingly complex which create cases where a mathematical model may not be easily formulated or implemented. Here, an algorithm can be used to build a machine by incorporating data which may be coupled with system models, regardless of their completeness. Currently, the use of multiple layers of ML algorithms to improve performance and broaden generalization, known as deep learning \cite{lecun2015deep}, is garnering the attention and focus of the research community. 

SDRs are flexible and can dynamically alter functionality across the protocol stack which creates numerous areas of research interests and capabilities. Recent prevalence of 4th generation wireless techniques at the PHY layer has redirected SDR research to multiple-input multiple output (MIMO) antenna systems and orthogonal frequency-division multiplexing (OFDM) applications. Similarly, SDRs are being used to investigate spectrum use optimization through congitive radio, dynamic spectrum access\cite{2014Kumar}, and software-defined networking at and above the MAC layer. In applications and complex concepts such as 5th generation wireless communications\cite{2015Wirth}, the flexibility and breadth of SDR approaches become ideal for low cost rapid prototyping.

By coupling the innovative capabilities of both SDR and ML, formerly impractical and computationally intensive tasks in communication systems are being accomplished on easily portable computational platforms. These couplings of SDR and ML result in a directional shift of modern communications systems such as in adaptive equalization \cite{widrow1975adaptive},\cite{ibnkahla2000applications}, spectrum sensing for cognitive radio\cite{bkassiny2013survey}, protocol identification\cite{hu2014mac}, and network optimization\cite{zorzi2015cognition}. Deep learning has been applied in Rayleigh fading channels for massive MIMO systems \cite{liao2018rayleigh}, been shown to synthesize modulation where the channel is previously unknown in adversarial networks \cite{o2018physical}, and has been applied as an autoencoder for OFDM schemes with non-linear amplifiers \cite{felix2018ofdm}. At the physical layer, ML is used in place of specific digital communications blocks such as: pulse-shaping filters, channel filters, serializers, forward-error correcting codes, and so on. Instead, network graphs are used to govern the operations of these blocks at the transmitter and receiver \cite{2017OSHEA,8054694,8214233}. The neural networks, or flow graphs\footnote{There is some ambiguity over the term ``network'' between the machine learning and communications communities. For clarity, ``network'', by itself, refers to the communications network while we use the word ``graph'' to refer to the learning network.}, are ``trained'' at their respective transmitter or receiver. After training, the learned machines, coupled with traditional communication blocks, are used for the PHY layer ML schemes.

In contrast to these discussed PHY techniques, this paper proposes the use of SDRs and ML in an attempt to replace {\it all} traditional signal processing blocks in order to create an expendable, yet resilient, communication scheme. A realistic framework is presented, referred to as deep modulation or deepmod, that can be used to generate a temporary PHY layer based on the channel and requirements of the communicating nodes. Upon the arrival of communication disruption or attack, such as a jammer, the graphs can be retrained allowing the nodes to recover communication. The link protocol is generated {\it ad hoc} and satisfies the current channel requirements. This protocol can be disposed of when: communication ends, periodically, or during an attack. As a result, two nodes create a unique communication chain that satisfies throughput requirements and can be discarded for a new and unique chain in the future. 

\begin{figure*}
  \includegraphics[width=\textwidth]{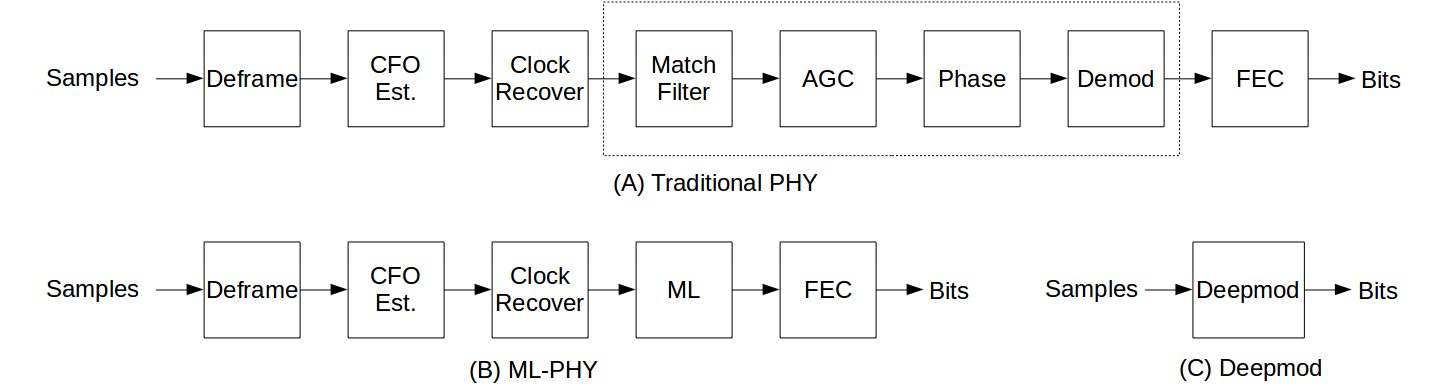}
  \caption{Example receiver processing blocks for A) a traditional PHY, B) a machine learning enabled PHY, and C) the proposed deepmod self-taught PHY.}
  \label{fig:system}
\end{figure*}

{\bf Novelty}: To the best of our knowledge, we show for the first time a single machine learner able to create a holistic, self-taught P2P PHY that can communicate over acoustic, powerline communications (PLC), and radio frequency (RF) channels - all without changing software. Holistic in that, at the transmitter, information bits are converted directly to samples\footnote{More ambiguity between communications and machine learning communities. ``Samples'' is used for those values generated by the SDR while ``examples'' is instead used for inputs into ML graphs.} by the machine (deepmod); and, at the receiver, samples are mapped to classes using deepmod and then converted to bit estimates. Rather than replacing some individual signal processing blocks, deepmod creates and learns the entire software portion of the PHY layer in a self-taught manner.

\section{System Model}

To scope the presentation of this work, this paper focuses on the point-to-point (P2P) link where both Users are enabled for full duplex communications. The P2P channel then consists of two Users each equipped with an appropriate transducer (e.g. antenna) for that channel. This transducer is used to convey energy representing information over an unknown channel
\begin{eqnarray}
    y_i & = & h_i\left(f_i(x_i)\right)+\eta_i
    \label{eq:system} \\
    \hat{x_i} & = & g_i(y_i)
    \label{eq:systemrx}
\end{eqnarray}
where $x_i$ are the information bits destined for User $i$. These bits are transformed to samples through PHY processing $f_i(\cdot)$ to then be transmitted over-the-air by the SDR. The channel $h_i(\cdot)$ corrupts the samples in some unknown manner and also includes adding $\eta_i$ as additive white Gaussian noise (AWGN) with variance $n^2_{std}$. For simplicity, \eqref{eq:system} assumes received samples $y_i$ are processed at baseband and signal magnitudes are bounded by unity. Estimates of information-bearing bits \eqref{eq:systemrx} are recovered through further processing, $g_i(\cdot)$, at the receive User.

\subsection{Traditional PHY}

The traditional approach to wireless communications is to find optimal bulk processing at the transmitter and receiver, $f_i(\cdot)$ and $g_i(\cdot)$, respectively, given some fixed distribution of the channel conditions $h_i(\cdot)$ and noise power $n^2_{std}$. Often, samples pass through a long chain of signal processing in order to correctly encode and decode the bits \cite{proakis1995}. Fig. \ref{fig:system}(A) shows what $g_i(\cdot)$ may look like at the receiver for a traditional approach to PHY layer processing. Raw samples are deframed according to some bursty traffic pattern. The carrier frequency offset (CFO) and timing mismatch, due to sensitive differences in User hardware, are recovered and corrected. Corrected samples are matched filtered and passed through an automatic gain controller (AGC) in case of multilevel modulation. After phase correction, the symbols are demodulated into bits and then corrected based on whatever forward error correction (FEC) was used at the transmitter. Though traditional PHYs may differ in order and types of processing from that shown in Fig. \ref{fig:system}(A); in general, each operation used on the data is ``optimal'' with regards to some metric such as bit-error rate (BER) or throughput for the given channel. Unfortunately, it is straightforward to alter $h_i(\cdot)$ or $\eta_i$ sufficiently, through an attack, an incidental interferer, or failure \cite{osti_1491329}, such that the optimal traditional processing also fails.

\subsection{Machine Learning PHY (ML-PHY)}

With the advent of SDR, many of these traditional signal processing blocks can be done digitally, and in real-time, after being sampled inside the radio. One use of ML in digital communications is to go further and remove these individual signal processing blocks altogether and simply pass the data through a ML graph, to perform, maybe not similar computations, but at least achieve similar results as suggested in \cite{2017OSHEA}. For example, consider the ML-PHY processing chain shown in Fig. \ref{fig:system}(B). Rather than using strict signal processing blocks, several of the blocks within the highlighted region are instead replaced by a machine that learns to do equivalent processing. This is immediately beneficially especially in channels where traditional blocks that were once optimal but become degraded through unknown channel effects or attacks.

Though using ML-PHY is interesting it is often not practical for two-way P2P links. In order to function correctly, the ML graphs must backpropagate calculations from the loss function through both graphs in order to update neuron weights. In a live system, this backpropagation cannot happen over the wireless channel in the traditional sense since there is not an instantaneous and error-free link between the source and sink Users. Additionally, there is no guarantee that the forward and reverse wireless channels are not heterogeneous. Finally, the above ML-PHY example would only replace a few of the many blocks required in a traditional processing channel. The following section attempts to design a more realizable, and holistic, framework for using ML in communications channels as shown in Fig. \ref{fig:system}(C).

\section{Deep Modulation (Deepmod)}
\begin{figure}
	\centering	
		\includegraphics[width=3.2 in]{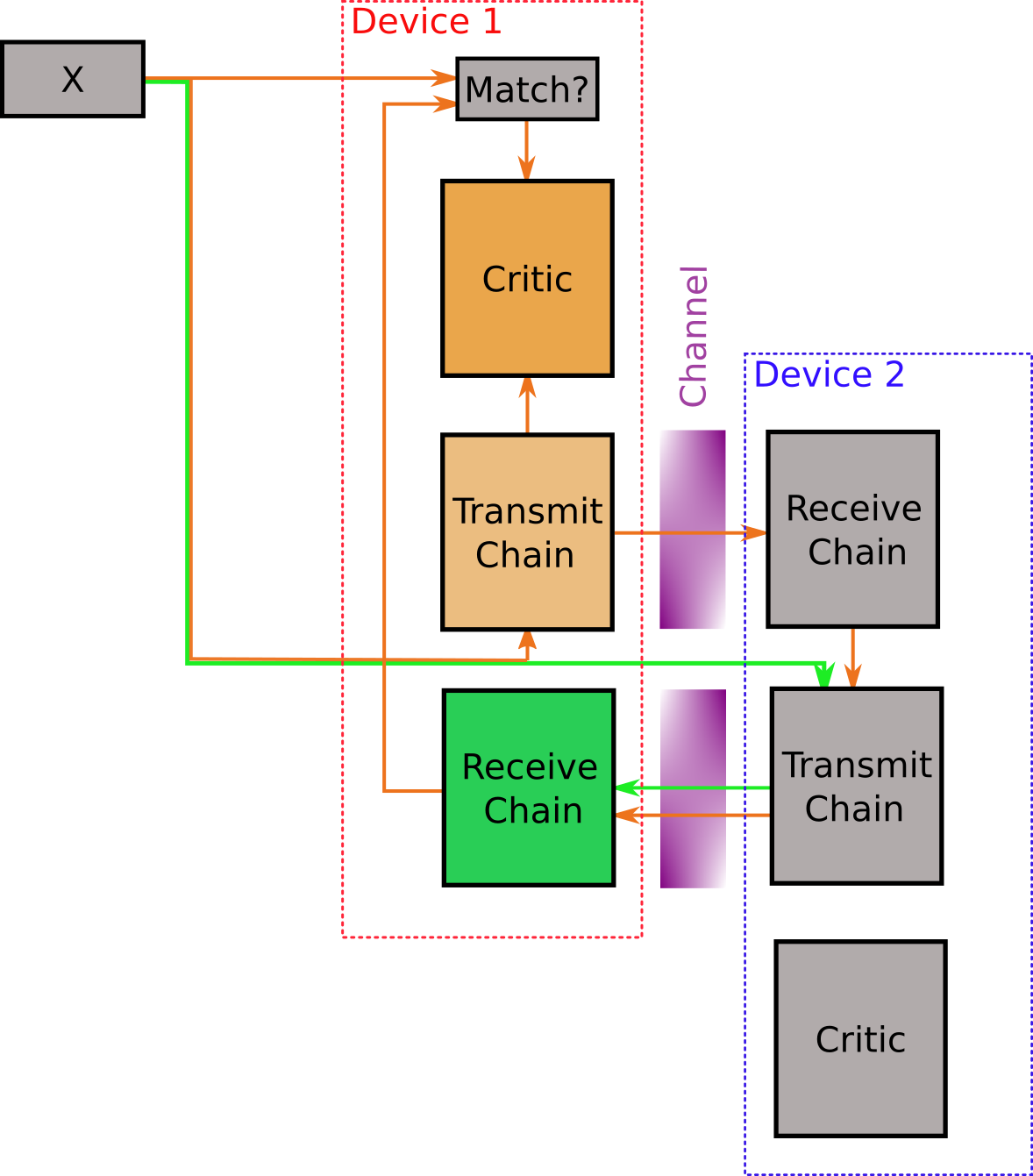}
    \caption{The adversarial deepmod PHY graph topology. A device can train its transmitter by utilizing a critic trained to recognize messages that will be successfully received by the other device (orange path). The receiver is only trained on whether or not it could properly decode the message from the known sequence (green path).}
		\label{fig:Deepmod2}
\end{figure}

The idea of deepmod is to 1) Replace all traditional system blocks and 2) Overcome some of the practical difficulties of using ML-PHY in digital communications. By definition, deepmod must be self-taught as it learns the PHY layer and User graphs must converge independently over the channel. Consider the system described in Fig. \ref{fig:Deepmod2} which is referred to as the deep modulation framework or just deepmod for short. The transmit and receive chains are suggestive of autoencoders with convolutional neural networks (CNN); however, additional learning blocks are introduced to solve the problem of unrealistic feedback channels and signal asymmetry. Both User 1 and User 2 have separate network modules for their transmit and receive chains; \hl{a} ``critic'' graph is added to the traditional autoencoder CNN. The purpose of this framework is so that backpropagation of the graph happens only on the User's own node and not across the channel while still allowing meaningful training to occur. 

Training happens in ``epochs'' where all classes are transmitted as a batch over the channel. The deepmod loss functions $L$ are designed to minimize the class error detection
\begin{eqnarray}
    L_{RX} &=& -\sum_{i=0}^{N-1} x_i \log(X^{\prime}_i)
    \label{eq:costrx} \\
    L_{TX} &=& -\log(C) \\
    L_{CRIT} &=& 
        \begin{cases}
            -\log(C), & \text{if}\ x = x^{\prime} \\
            -\log(1-C), & \text{otherwise} \\
        \end{cases}
    \label{eq:cost}
\end{eqnarray}
where $N$ is the total number of classes. From \eqref{eq:system}, information bits are mapped to classes where a single time instance of the $i$th class is $X_i$ (a one-hot vector). This class is encoded as $Y_i$ (a group of samples or ``waveform'') by the transmit deepmod chain which passes through the wireless channel and is decoded at the sink User. The sink User then re-transmits the re-encoded message back through its own channel and the source User's receive chain decodes the message as $X_i^{\prime}$ as in \eqref{eq:costrx}. The critic is trained to predict value $C$ which indicates whether or not the sink User will be able to properly decode all $Y_i$. Figure \ref{fig:Deepmod2} demonstrates how the first path is used to train the critic and the transmit chain, and the second path is used to train the receive chain - all distributed and ``over-the-air''.

\hl{A detailed explanation of the machine learning graphs may be useful to understand how the system is able to train. Apart from the additional critic graph, deepmod follows a typical approach to deep autoencoding CNNs but applied to digital communications. At the transmitter, after bits are mapped to one-hot class labels, they are input into the encoder. An embedding layer maps the total number of input classes to a smaller embedding size. This is followed by two fully connected layers with $tanh$ activation functions. The output layer size is equal to the number of samples per class. Due to the activation functions, these sample magnitudes are bounded by unity and transmitted over the channel medium assisted by a digital-to-analog converter (DAC) and hardware amplification. At the receiver, complex samples, taken directly from the analog-to-digital converter (ADC), are deinterleaved into real and imaginary parts and input as features into the decoder. The decoder starts with two fully connected layers, which shrink the input feature size, followed by a convolutional layer. Max pooling is then used before a final linear layer to output decoded class probabilities.}

This approach to a self-taught PHY layer in digital communications is similar to approaches taken in generative adversarial networks \cite{goodfellow2014generative} where, instead of having a known set of ``real'' classes, deepmod nodes ``compete'' in developing their own PHY language classes understood by both Users. The learned weights of each portion of the network (i.e. receive chain, transmit chain, and critic) are updated based only on their corresponding loss function. The receive chain is only updated based on the known sequence, and not on the decoded/re-encoded message from the other User. 

\section{Results}
The purpose of this paper is to show, experimentally, how deepmod can learn to exchange information over a variety of channel media without using any traditional processing blocks. All forward and reverse processing from \eqref{eq:system} must be learned and then implemented by deepmod. The following experiments were run using TensorFlow \cite{tensorflow2015} and Gnuradio \cite{gnu-radio} on a deepmod machine modem as described above. TensorFlow creates a fully connected graph when generating the gradient; to ensure that backpropagation only occurs within a User's own graph, care must be taken that backflow is stopped from the other User. For these experimental results, deepmod graphs are distributed on two separate laptop computers with no connections besides the wireless medium ensuring that no backpropagation occurs directly across the channel.

\subsection{Channel Media}

\begin{table}[]
\caption{Parameters for Different Channel Media}
\label{table:parm}
\begin{tabular}{@{}llllll@{}}
\toprule
Medium    & fc (Hz)  & Sample Rate & Transducer  & USRP & PHY     \\ \midrule
RF        & 900M & 1 Msps      & VERT900    & B210 & deepmod \\ \midrule
Powerline & 83M  & 1 Msps      & Coupler    & B210 & deepmod \\ \midrule
Acoustic  & 0    & 44.1 ksps   & Speaker/Mic & N210 & deepmod \\ \bottomrule
\end{tabular}
\end{table}

In \cite{osti_1491329} favorable simulated and experimental results encourage closer examination into the concept of deepmod in a wider variety of channels. Details of these different media and hardware are explained below; however, Table \ref{table:parm} details the key hardware and parameter settings used for the channels. The primary takeaway from this table is that all media use the same PHY layer learner - deepmod - but with vastly different hardware and settings. For example, a 900 MHz signal, as in the RF channel, propagates fine in freespace with a VERT900 antenna but does not propagate well in the PLC channel and would not propagate at all in the extremely low frequency acoustic channel with a speaker and microphone combo; however, as shown below, deepmod learns a PHY protocol regardless of this extreme difference and viable communications takes place without traditional communications assistance. 

\subsubsection{RF Channel}

\begin{figure}
	\centering	
		\includegraphics[width=3.2 in]{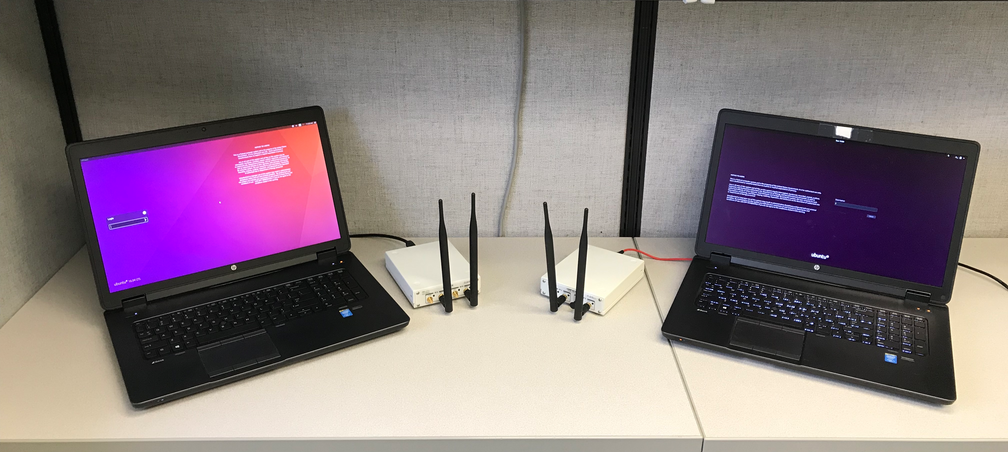}
    \caption{RF channel experimental hardware for deepmod including B210 USRPs and VERT900 antennas.}
		\label{fig:dm_rf}
\end{figure}

The RF channel represents the typical wireless environment such as that seen in the ISM band. For the RF experiments, we use Ettus Research's B210 SDR universal software radio peripheral (USRP) transceiver. These radios can operate at a center frequency from 70 MHz to 6 GHz with an effective bandwidth up to 56~MHz. Though the software-defined nature of this setup allows us to experiment in a variety of channel conditions by simply altering a few input variables such as center frequency and sample rate for purposes of these RF experiments the center frequency is fixed at 900~MHz, radios are equipped with VERT900 antennas, and the sample rate is set to 1 Msps. The transmit gain is adjusted, and received SNR calculated, to demonstrate deepmod performance over a wide range of receive SNR values for both training and testing. The experimental hardware setup for the RF channel is shown in Fig. \ref{fig:dm_rf}.

\subsubsection{Powerline Channel}

\begin{figure}
	\centering	
		\includegraphics[width=3.2 in]{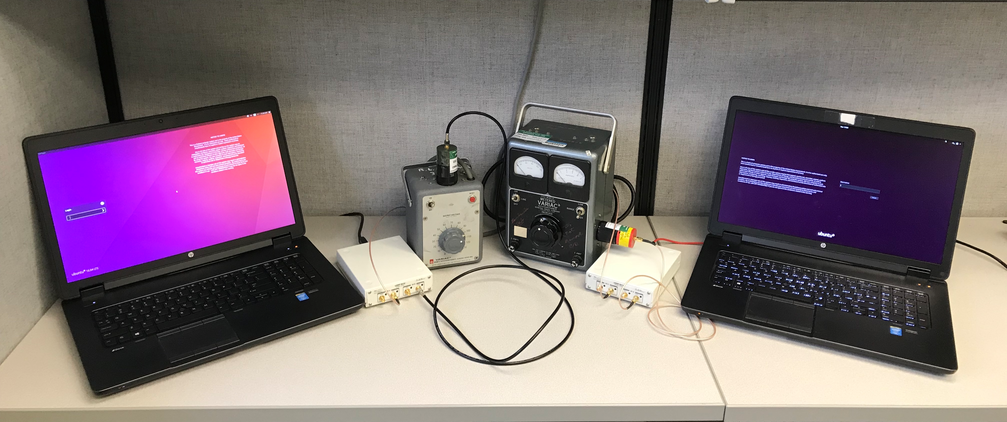}
    \caption{PLC channel experimental hardware setup for deepmod with high-pass filters protecting the B210s from the 60~Hz line power and variacs to simulate additional microgrid transformers.}
		\label{fig:dm_plc}
\end{figure}

The PLC channel, though not as prolific as RF, has seen considerable use in commercial and academic enterprises \cite{facina2016cooperative}. For purposes of this benchtop experiment, a custom high-pass filter coupler was designed to ensure the USRPs were not damaged when operating over the high powerline voltages. To enable some customization of the channel, two 115 V input, 0-135 V output variacs are included in the forward and reverse links giving some control to the PLC channel characteristics. The experimental hardware setup for the PLC channel is shown in Fig. \ref{fig:dm_plc}.

\begin{figure}
	\centering	
		\includegraphics[width=3.2 in]{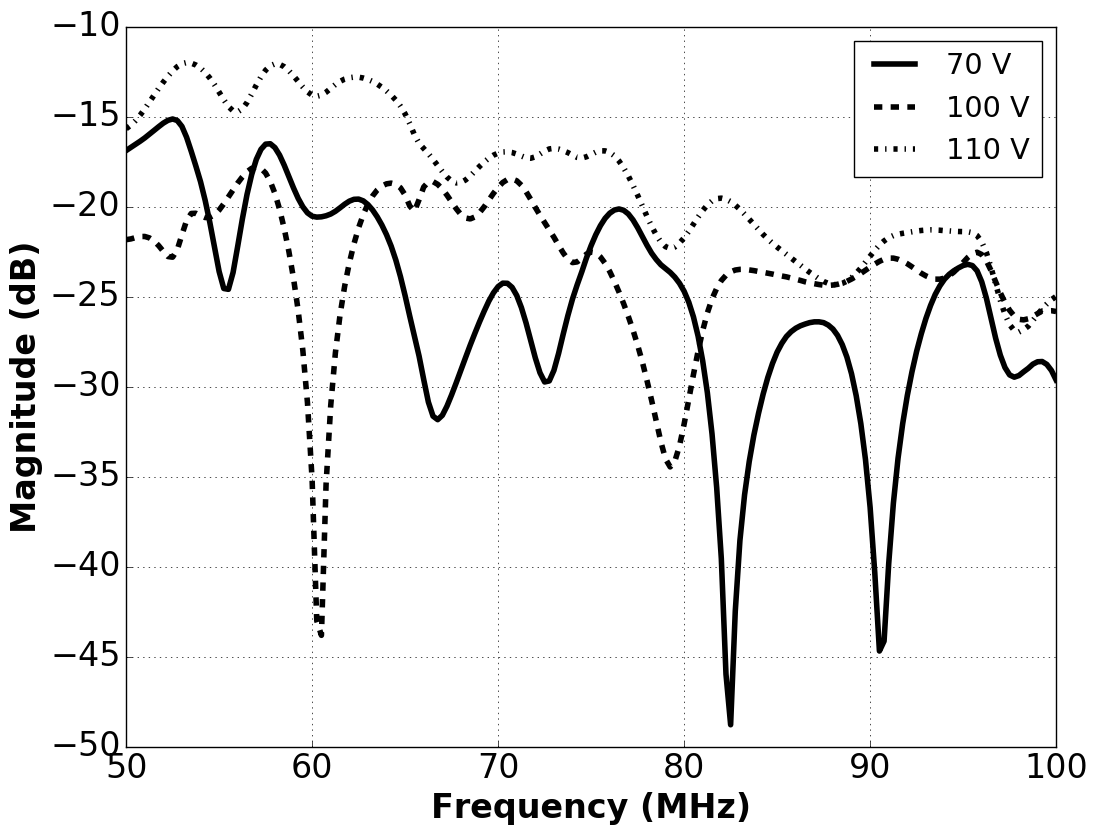}
    \caption{Frequency magnitude response of the PLC channel from 50 to 100~MHz for various output voltages of the variac transformer.}
		\label{fig:dm_mag}
\end{figure}

It is well known that PLC channels have potentially rapid variations in the frequency magnitude response \cite{liu2014power}. Traditional communication protocols often assume so-called ``flat fading'' and rely on schemes such as OFDM when the channel is more selective. For example, the RF channel used in this work would be considered frequency flat. Ideally, deepmod is able to adapt and learn in whichever channel it is placed in regardless of the response characteristics. Fig. \ref{fig:dm_mag} shows the response for our benchtop microgrid hardware for a given turns ratio on the General Radio W5MT3 Variac. For testing purposes, we intentionally choose a center frequency such that the channel becomes frequency selective over the 1 MHz bandwidth. This is done to contrast the PLC channel with the frequency-flat RF channel and better demonstrate the power of using deepmod in the PHY layer.

\subsubsection{Acoustic Channel}

\begin{figure}
	\centering	
		\includegraphics[width=3.2 in]{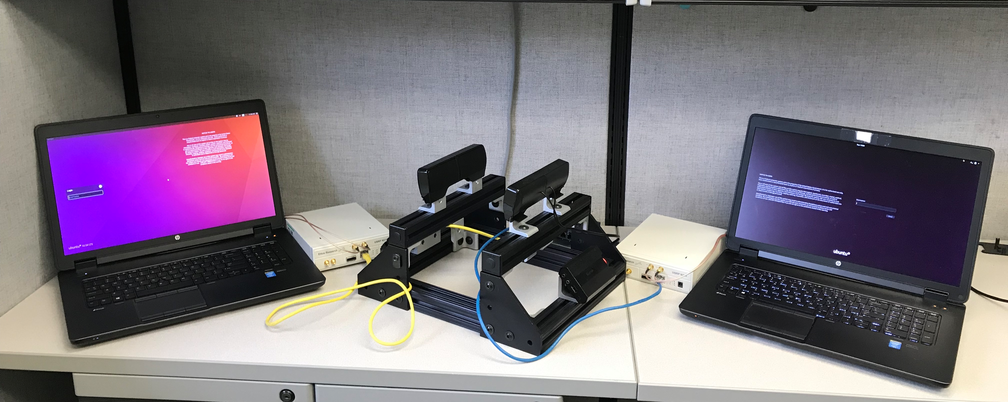}
    \caption{Acoustic channel experimental hardware setup: N210 USRPs with LFRX/TX daughtercards, network switch, and speakers/microphones for signal transduction.}
		\label{fig:dm_acoustic}
\end{figure}

The acoustic channel is used for both over-the-air transmissions as well as underwater communications \cite{barbeau2017weak}. As the acoustic channel uses extremely low frequencies without RF energy, a different set of transducer hardware is required for this channel. The USRPs are changed to N210s with LFRX/TX daughtercards which can transmit at baseband with no carrier frequency. The trade-off is these daughtercards have no modifiable gain functionality disallowing power sweeps in the experimental results. The speaker and microphone are generic off-the-shelf models used in typical home computer environments. For the acoustic experiment, in addition to swapping transducers, the sample rate of the USRPs is set to 44.1 kHz, which is much smaller than that of the other higher frequency RF and PLC channels, and samples are transmitted at baseband (no carrier frequency). The experimental hardware setup for the acoustic channel is shown in Fig. \ref{fig:dm_acoustic}.

\subsection{Deepmod Convergence}

When including any ML operations in communications networks \hl{there is a natural question that arises: How long does training take to converge? To answer this question,} over-the-air experiments were run using deepmod in the RF, PLC, and acoustic channels. For each experiment, the parameters from Table \ref{table:parm} were used. For training, deepmod initializes 256 random waveforms representing 256 classes. One training epoch consists of all possible classes sent across the chosen channel. Subsequent epochs contain a stochastic permutation, for training robustness, which is a typical approach to machine training. Each epoch is transmitted with a small time delay between batches so that frames can easily be detected by the USRP.

\begin{figure}
	\centering	
		\includegraphics[width=3.2 in]{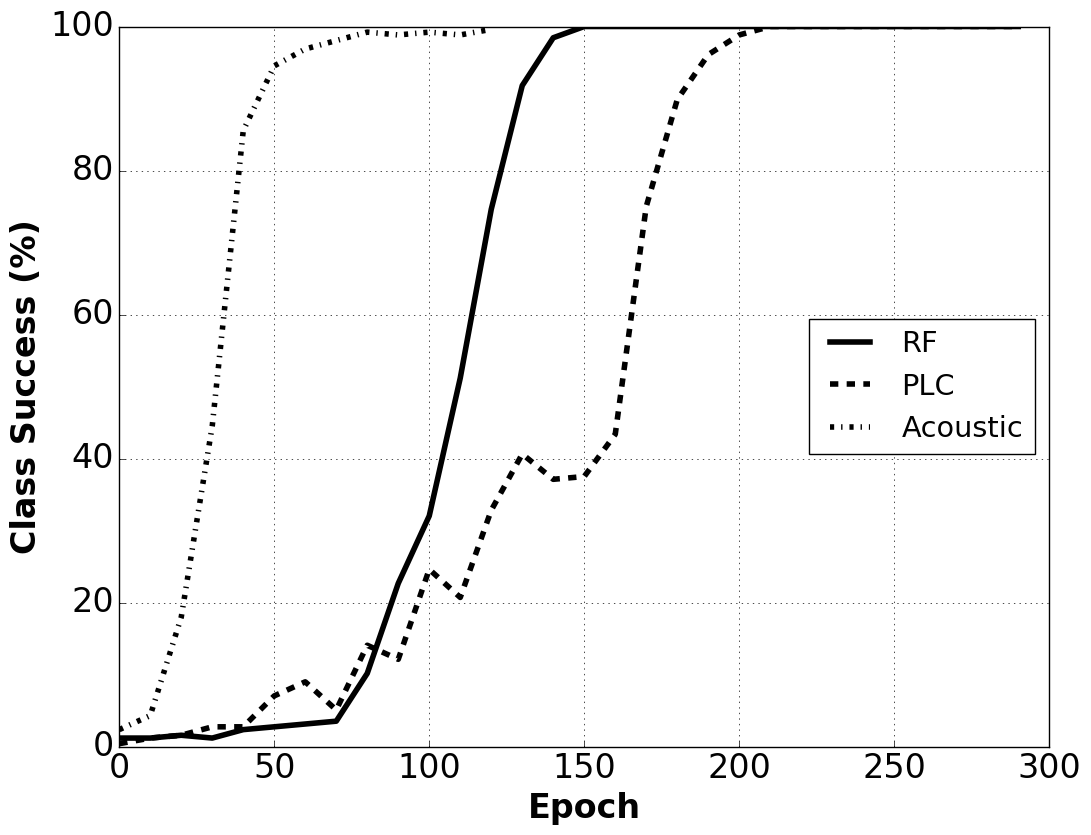}
    \caption{Class success percentage of deepmod training versus the number of epochs trained. Different curves represent training time in different channel media. Train SNR is fixed at roughly 10 dB.}
		\label{fig:convergence}
\end{figure}

Fig. \ref{fig:convergence} attempts to answer the question of deepmod convergence. Each curve represents the percent total number of correctly decoded classes per epoch as a function of epoch count. Initially, with random initialization of classes, the Users are unable to successfully decode each other's language. As time progresses, each channel successfully converges to a high probability of decoding most classes - all in under 200 epochs; however, the behavior of learning is quite different. The convergence time seems to follow the complexity of the channel involved which makes sense from a learning perspective. Where the RF and acoustic channels resemble AWGN, the RF channel is much wider in frequency and takes longer to learn than the acoustic channel. The PLC channel is the same bandwidth as the RF channel but is frequency selective resulting in greater training difficulty as shown by the jagged curve behavior and longer convergence time.

The number of epochs for convergence must be converted to seconds for a better comparison. The time to convergence, rather than number of epochs, depends on system parameters used such as sample rate and samples per class. For example, for batches containing all possible input classes, deepmod requires 2048 samples per epoch when using eight samples per class. In a 1 MHz RF channel - radios set at 1 Msps - the roughly 200 epoch training period can be accomplished realistically in just a few seconds even with the padding placed between epochs. The acoustic channel, with a faster convergence time in epochs, actually takes longer to train due to the lower sample rate of 44.1 \hl{kHz}.

\subsection{Deepmod Performance}

Similar to the deepmod convergence test, performance can be analyzed by first training at a certain power (called train SNR) and then sweeping test SNR in the different channels. It should be noted that these performance curves are class-error rate and not the traditional bit-error rate. 256 classes are used for the deepmod graphs with 16 real (or eight complex) samples at the inner layer. These values are 8 bits per class, and with transmitters set to 8 samples per class, result in a spectral efficiency of 1 bit per sample (potentially 1 Mbps for the given sample rate) of uncoded throughput. Uncoded since deepmod was not tasked with learning error correction for these experiments.

\begin{figure}
	\centering	
		\includegraphics[width=3.2 in]{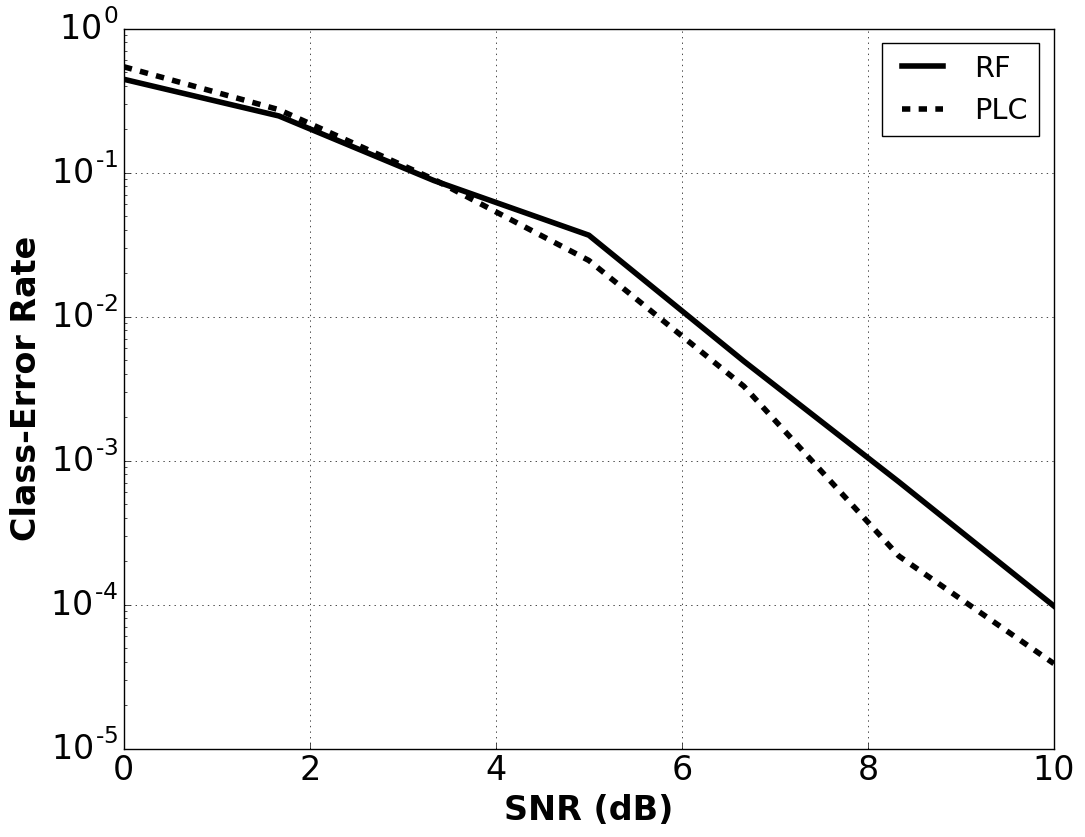}
    \caption{Experimental class-error rate (CER) of deepmod in the RF and PLC channels. Train SNR is fixed at 10 dB while test SNR is swept as shown.}
		\label{fig:cer}
\end{figure}

Fig. \ref{fig:cer} shows the class-error rate (CER) of the RF and PLC channels when each are trained at 10 dB SNR and then swept test SNR as shown. The acoustic channel CER results are omitted as the hardware difference (no daughtercard gain) precluded a fair, and repeatable, comparison in received SNR values; however, Fig. \ref{fig:convergence} already showed that the acoustic channel can converge to a good CER performance. \st{For the given spectral efficiency, the CER performance of RF and PLC is on par with traditional communications of similar spectral efficiency and channels.}

It should be noted that the curves in Fig. \ref{fig:cer} refer to measured receive SNR and not transmit power. The pathloss in the PLC channel, with the chosen variac settings, is much greater than the RF channel for the distances between antennas. The USRP gain values are adjusted to compensate so that a comparison between received SNR can be made. Therefore, the difference in the curves is due to the nature of experimental results and not necessarily channel effects.

\st{In addition to showing how deepmod can replace traditional PHY layer processing, } \hl{This} experiment emphasizes another primary attribute of deepmod - its resilience or ability to change channel media on-the-fly without modifying PHY layer code. The exact same deepmod enabled laptops used for the RF experiment are used in the PLC channel simply by changing the transducers (antennas to couplers) and gain values. For example, one might envision next-generation smart grids utilizing PLC channels for data backhaul and secure communications. A catastrophic event or attack which renders PLC unusable, or severely limited, could be diverted immediately to RF simply by deepmod relearning to communicate in the new medium or remaining in PLC and learning around the event.

\subsection{A Note on Learning}

Though an exhaustive discussion on the implications of deepmod for digital communications is beyond the scope of a single paper, it is worthwhile to close with \hl{some final notes on how the machine learns to communicate, specifically, regarding hyperparameters used in deepmod neural networks and how deepmod is able to learn digital communications}. 

\begin{figure}
	\centering	
		\includegraphics[width=3.2 in]{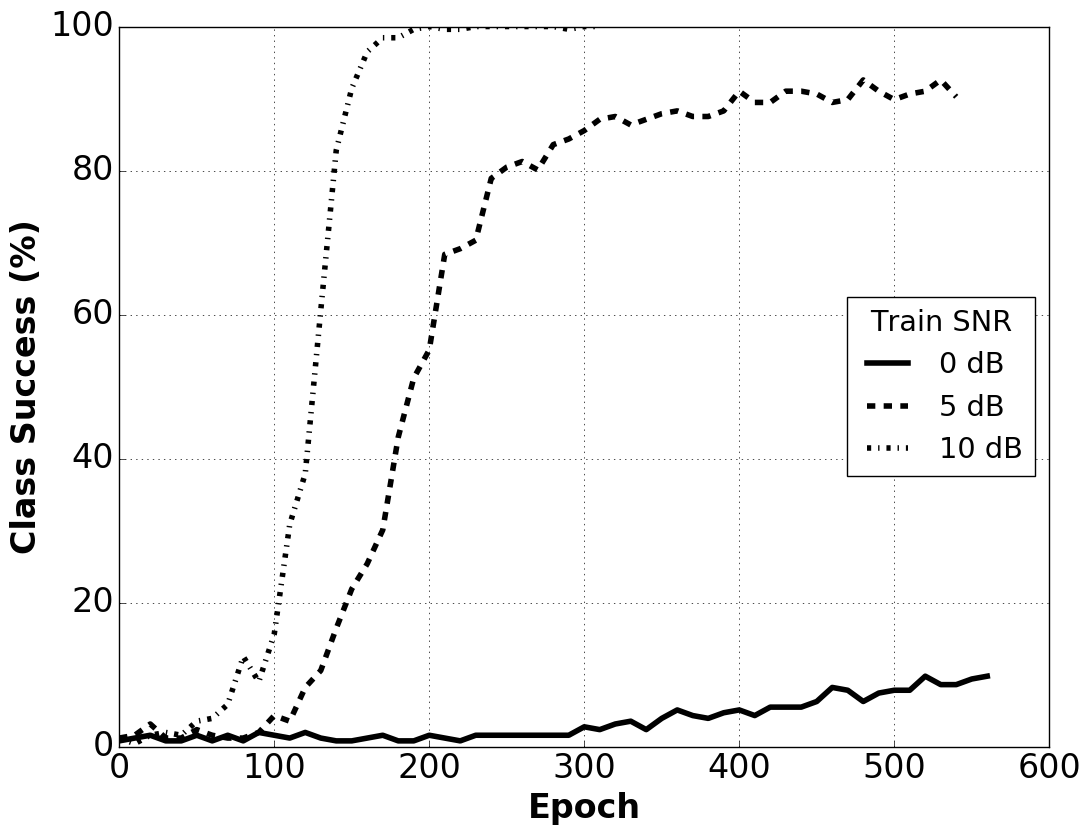}
    \caption{Class success percentage of deepmod training versus the number of epochs trained in the RF channel. Different curves represent the different train SNRs.}
		\label{fig:convergence_db}
\end{figure}

One of the many hyperparameters in the deepmod graph is training SNR level. A popular question may \hl{be: What's} the optimal power level for training in a certain environment? To consider this question, two experiments were run in the RF channel where deepmod was trained at certain SNR levels and then tested at a different set of SNRs. First consider the convergence behavior of deepmod in the RF channel as shown in Fig. \ref{fig:convergence_db}. For these results the environment is fixed at RF while the train SNR is swept from 0 to 10 dB for various test SNR values. A valid conclusion is that the training time required for convergence decreases with increased test SNR. It is also shown that the system converges to the CER performance based on the train SNR value; however, this does not answer the question of CER performance as a function of train SNR.

\begin{figure}
	\centering	
		\includegraphics[width=3.2 in]{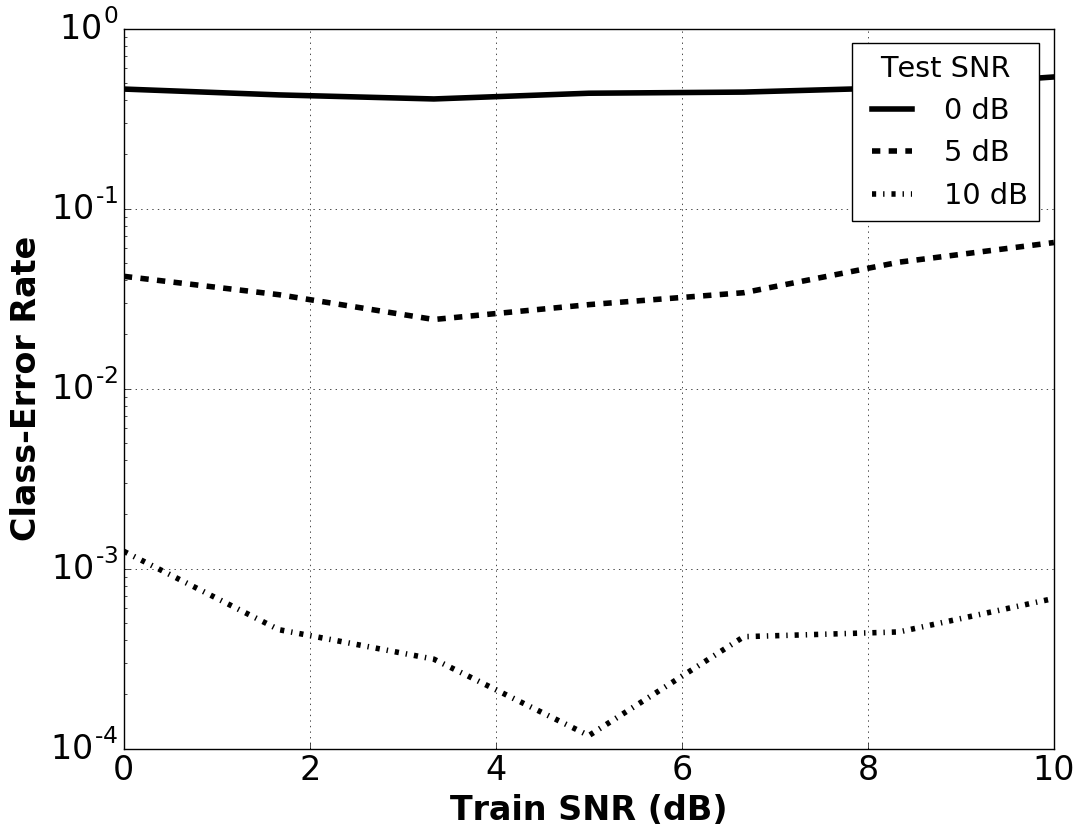}
    \caption{CER performance at various test SNR levels as a function of the deepmod train SNR in the RF channel. Note each curve is roughly convex.}
		\label{fig:train_snr}
\end{figure}

The results of this second experiment are shown in Fig. \ref{fig:train_snr} where a cross-section of the CER performance curve is shown for swept train SNR. It may be surprising to note that the performance of each test SNR is roughly convex as a function of the training SNR rather than strictly increasing or decreasing. Training at maximum SNR is not necessarily the optimal hyperparameter setting. This is easier to understand by considering the two edge cases: no noise (infinite train SNR) or no signal (zero train SNR). With no signal, the machine obviously cannot learn anything about the PHY layer and performance will be poor; however, a similar phenomenon occurs with no noise. At infinite SNR, the waveforms (representing classes) imagined by deepmod converge quickly to a solution that satisfy the machine's cost functions. Limited noise is too easy on the machine. The best learning takes place when deepmod must work at producing waveforms that function in noise - even if such learning never converges to perfection. This idea is not too dissimilar from data augmentation \cite{wong2016understanding} in computer vision and image machine learning algorithms where training images are intentionally distorted by a variety of operations (cropping, shearing, rotating, etc.) to improve machine learnability.

\hl{Finally, at first glance, it seems incredible that deepmod is able to relearn transmit and receive processing chains on its own. However, the CNN that defines deepmod is equipped with all the functionality required to reproduce traditional signal processing blocks. For example, adaptive AGCs are often used in software for multi-level modulations such as QAM. This is done in traditional schemes so that symbol magnitudes are bounded by unity for detection purposes. Analogously, deepmod contains $tanh$ activation functions, whose magnitude outputs are bounded by unity, that can result in similar behavior if training dictates such. Fully connected layers within the CNN have similar mathematical operations to traditional filters where weights are learned rather than preset as in matched filters. The comparisons can go on; however, the key concern is that traditional digital communication systems use human optimized blocks to find $f_i(\cdot)$ from (1) to convert bits to samples and  $g_i(\cdot)$ from (2), to convert samples to bits. When deepmod is given sufficient depth in its CNN it can simply learn the functions $f_i(\cdot)$ and $g_i(\cdot)$, on its own, through the training methodology described in this work.}

\section{Conclusion}
Deep modulation, or deepmod, is a machine learning framework designed to replace much of the traditional signal processing blocks at the PHY layer by creating a machine that is self-taught in how to exchange information over an unknown channel. Deepmod-enabled Users initialize with a set of unique classes (waveforms) to transmit data over the current channel medium. With the assistance of a specially designed critic graph, these waveforms converge quickly to a set of decodeable classes at the receiver. It was shown experimentally that deepmod can be used to successfully transmit bit-bearing classes / waveforms across an unknown channel \st{and provide performance comparable to some traditional systems} - even across different media such as RF, acoustic\st{s}, or PLC channels. These media represent radically different environments from frequency-flat to frequency-selective channels as well as narrow- and wideband spectrum usage. Deepmod then has an inherent attribute of resilience as well as adaptability for just-in-time communications as the same machine can learn to communicate over different channels given the appropriate hardware transducer. 

\bibliographystyle{IEEEtran}
\bibliography{bib/refs}

\end{document}